%% file: WSchlueter_2.tex
\documentclass{evn2004}
\usepackage{txfonts}
\usepackage{graphicx}
\begin{document}
\include{page}
   \title{IVS Products for Precise Global Reference Frames}

   \author{Wolfgang Schl\"uter\inst{1}
          \and 
          Nancy Vandenberg\inst{2}                                          
          }

   \institute{Bundesamt f\"ur Kartographie und Geodaesie, Fundamentalstation Wettzell, D 
              93444 K\"otzting, Germany
              \and
              NVI, Inc./ Goddard Space Flight Center, Greenbelt, USA   
             }

   \abstract{ VLBI plays a unique and fundamental role in the
    maintenance of global reference frames which are required for
    precise positioning in many research areas related e.g to the
    understanding and monitoring of global changes, to geodesy, to
    space missions etc. The International VLBI Service for Geodesy and
    Astrometry coordinates internationally the VLBI components and
    resources and is tasked by IAG and IAU for the provision of
    products describing the Celestial Reference Frame through positions
    of quasars and their changes with time and also for products
    describing the rotation of the Earth in space. This paper
    summarises today's status of the products, achieved over the last
    years by evolving the observing programs. It points out the
    activities to improve further on the product quality to meet future
    service requirements, which will come up with the need for highly
    precise global reference frames.  }

   \maketitle
%

\section{Introduction}

The International VLBI Service for Geodesy and Astrometry (IVS) is a
service of the International Association of Geodesy (IAG), the
International Astronomical Union (IAU) and the Federation of
Astronomical and Geophysical Data Analysis Services (FAGS).  The main
task of the IVS is the coordination of VLBI components in order to
guarantee the provision of the products and parameters to realize the
Celestial Reference Frame (CRF), the Terrestrial Reference Frame (TRF)
and the Earth Rotation with its orientation of the rotation axis in
both reference frames and its angular velocity through the Earth
Orientation Parameter (EOP).  The EOP enables the transformation
between TRF and CRF.  VLBI is fundamental and unique for the
realization of the CRF through a catalogue of quasar positions.  The
IAU tasked IVS to maintain the CRF and released a resolution in August
2001, during the General Assembly in Birmingham. VLBI contributes
strongly to the TRF by the determination of station positions, in
particular of baseline lengths between the stations.  Due to the long
intercontinental baselines VLBI is strongly supporting the scale of the
TRF.  Time series of station positions give information about their
movements (plate motions).  As VLBI provides the complete set of EOP
and uniquely the UT1-UTC (DUT1) parameter and the CRF, VLBI is the key
technique for the monitoring of global reference frames. The
international collaboration is based on a Call for Participation in
1998 with respect to the IVS Terms of Reference (ToR). Proposals were
made to operate more than 73 permanent components by 37 institutions in
17 countries. IVS has more than 250 Associate Members. Annual Reports
and Meeting Proceedings are published (Vandenberg \cite{vb99},
Vandenberg, Baver \cite{vb01}, Vandenberg, Baver \cite{vb02} ,
Vandenberg, Baver \cite{vb03} , Vandenberg, Baver \cite{vb04} ,
Vandenberg, Baver \cite{vbb00}, Vandenberg, Baver \cite{vbb02} ,
Vandenberg, Baver \cite{vbb04}).


\section{Products and improvements}

 \begin{figure*}
 \centering
  \includegraphics[angle=-90, width=\textwidth]{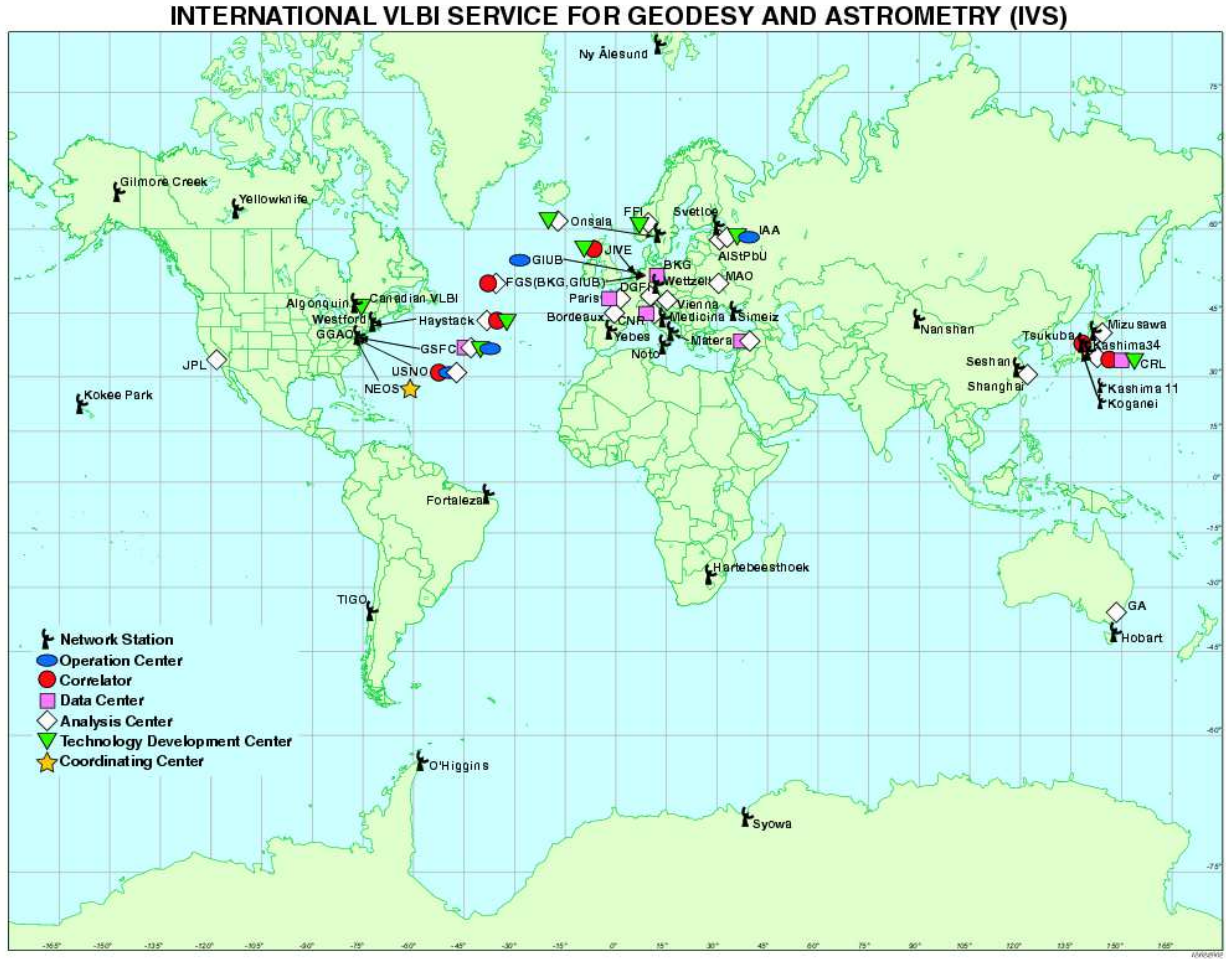}
  \caption{Overview of IVS components}
 \label{fig:ivs}
 \end{figure*}
 
When IVS started in March 1999, the demand for continuity in
maintaining the reference frames forced us to employ the existing
observing programs set up by the US Naval Observatory as NEOS, or by
NASA as CORE. In 2001 a working group (IVS WG2) was established to
review the products and the existing observing programs. The WG2 report
was the basis for improving products and evolving observing programs to
meet service requirements. The IVS products can be defined in terms of
their accuracy, reliability, frequency of observing sessions, temporal
resolution of the estimated parameters, time delay from observing to
final product, and frequency of solutions. The situation before 2002
and the goals for the follow on years with IVS products are described
in detailed tables in the WG2 report (Schuh et al. \cite{schuh02}).
The main IVS products, their current accuracies and the goals are
summarized in table 1. The VLBI technique allow us to provide
additional products and IVS intends to set up the extended products
summarized in table 2.

As of late 2001, IVS products were generated from ~3 days/week
observing with 6-station networks. The time delay ranged from several
days up to 4 months, with an overall average value of 60 days. Over the
next years, the goals of IVS with respect to its products were the
following (specific goals for each product are listed in the WG2 report
tables):

\begin{itemize}
\item improve the accuracies of all EOP and TRF products by a factor of
      2 to 4 and improve the sky distribution of the CRF,
\item decrease the average time delay from 60 to 30 days, and designate
      2 days per week as rapid turnaround sessions with a maximum delay
      of 3-4 days,
\item increase the frequency of observing sessions from 3 to ~7 days per week,
\item deliver all products on a regular, timely schedule.
\end{itemize}

\begin{center}
   \begin{table*}
      \caption[]{Summary of IVS main products, status and goal specifications}
         \begin{tabular}{|l|l|l|ll|}
            \hline
            \multicolumn{1}{c}{\bf Products} & \multicolumn{1}{c}{\bf Specifications} & \multicolumn{1}{c}{\bf Status} & \multicolumn{2}{c}{\bf Goals (2002 -- 2005)} \\
            \hline
            Polar Motion & accuracy              & $x_{p}$ \verb|~| $100\mu as$, $y_{p}$ \verb|~| $200\mu as$ & $x_{p}$, $y_{p}$: $50\mu as$ & \ldots $25\mu as$ \\
            $x_p$, $y_p$ & latency               & 1 -- 4 weeks \ldots 4 months                               & 4 -- 3 days                         & \ldots 1 day \\
                         & resolution            & 1 day                                                      & 1 day \ldots 1 h                   & \ldots 10 min \\
                         & frequency of solution & +3 days/week                                               &                                     & \ldots 7 days/week \\
            \hline
            UT1 - UTC    & accuracy              & $5\mu s$ \ldots $20\mu s$                                  & $3\mu s$ \ldots $2\mu s$            & \\
            DUT1         & latency               & 1 week                                                     & 4 -- 3 days                         & \ldots \ 1 day \\
                         & resolution            & 1 day                                                      & 1 day                               & \ldots 10 min \\
            \hline
           Celestial Pole& accuracy              & $100\mu as$ \ldots $400\mu as$                             & $50\mu as$                          & \ldots $25\mu as$ \\
$d\epsilon$, $\delta\psi$ & latency              & 1 -- 4 weeks \ldots 4 months                               & 4 -- 3 days                         & \ldots \ 1 day \\
                         & resolution            & 1 day                                                      & 1 day                               &  \\ 
                         & frequency of solution & \verb|~| 3 days/week                                       &                                     & \ldots 7 days/week \\
            \hline 
            TRF(x,y,z)   & accuracy              & 5 mm -- 20 mm                                              & 5mm                                 & \ldots 2 mm\\
            \hline
            CRF          & accurary              & $0.2\mu as$ -- $3 mas$                                     &  $0.25\mu as^{\mathrm{1}}$    & \\
            $(\alpha; \delta)$ & frequency of solution& 1 year                                                & 1 year                              & \\
                         & latency               & 3 -- 6 months                                              & 3 months                            & \ldots 1 month \\
            \hline  
         \end{tabular}
     \begin{list}{}{}
     \item[$^{\mathrm{1}}$] improved distribution
     \end{list}
   \end{table*}
\end{center}
%

\begin{center}
   \begin{table*}
      \caption[]{Extended products derived by VLBI and provided or intended to be provided by IVS}
         \begin{tabular}{|l|l|}
            \hline
            Earth Orientation Parameter additions & $\bullet$ dUT1/dt (length of day) \\
                                                  & $\bullet$ $dx_{p}/dt$; $dy_{p}/dt$\ (pole rates) \\
            \hline  
            Terrestrial Reference Frame (TRF)     & $\bullet$ x-, y-, z- time series \\
                                                  & $\bullet$ Episodic events \\
                                                  & $\bullet$ Annual solutions \\
                                                  & $\bullet$ Non linear changes \\
            \hline
            Celestial Reference Frame (CRF)       & $\bullet$ Source structure \\
                                                  & $\bullet$ Flux density \\
            \hline
            Geodynamical Parameter                & $\bullet$ Solid Earth tides (Love numbers h,l) \\
                                                  & $\bullet$ Ocean Loading (amplitudes and phases\ $A_{i}, \Phi_{i}$) \\
                                                  & $\bullet$ Atmospheric loading (site-dependent coefficients) \\
            \hline
            Physical Parameter                    & $\bullet$ Tropospheric parameters \\
                                                  & $\bullet$ Ionospheric mapping \\
                                                  & $\bullet$ Light deflection parameter \ $\gamma$ \\
            \hline  
         \end{tabular}
   \end{table*}
\end{center}
\section{Evolving observing programs }

To meet its product goals, beginning with the 2002 observing year, IVS
designed an observing program coordinated with the international
community. The 2002 observing program included the following sessions:
\begin{itemize}
\item EOP: Two rapid turnaround sessions each week (IVS R1 and IVS R2),
initially with 6 stations, increasing to 8.  These networks were
designed with the goal of having comparable\ $x_p$\ and\ $y_p$\
results. One-baseline 1-hr INTENSIVE sessions four times per week, with
at least one parallel session.
\item TRF: Monthly TRF sessions with 8 stations including a core
network of 4 to 5 stations and using all other stations three to four
times per year.
\item CRF: Bi-monthly RDV sessions using the Very Long Baseline Array
(VLBA) and 10 geodetic stations, plus quarterly astrometric sessions to
observe mostly southern sky sources.
\item Monthly R\&D sessions to investigate instrumental effects,
research the network offset problem, and study ways for technique and
product improvement.
\item Annual, or semi-annual if resources are available, 14-day
continuous sessions to demonstrate the best results that VLBI can
offer, aiming for the highest sustained accuracy.
 \end{itemize}
 
Although certain sessions have primary goals, such as CRF, all sessions
are scheduled so that they contribute to all geodetic and astrometric
products. Sessions in the observing program that are recorded and
correlated using S2 or K4 technology will have the same accuracy and
timeliness goals as those using Mark 4/Mark 5. The new IVS observing
program began in January 2002. The observing program and product
delivery was accomplished by making some changes and improvements in
IVS observing program resources (station days, correlator time, and
magnetic media), by improving and strengthening analysis procedures,
and by a vigorous technology development program.

\section{Status and experiences  of the new IVS Program two year after its implementation} 

The number of station observing days increased by about 10
compared to 2001, with an additional 12
campaign. Not counting CONT02, the number of observing days increased
by another 12
will continuously increase such that by 2005 the top dozen geodetic
stations will need to be observing up to 4 days per week - an ambitious
goal.  Increased station reliability and unattended operations can
improve temporal coverage by VLBI and also allow substantial savings in
operating costs.  Higher data rate sessions can yield more accurate
results, and therefore nearly all geodetic stations were upgraded to
Mark 5 and K5 technology. A deployment plan for the Mark 5 system was
proposed. As of the end of 2003 the correlator and most of the
observing stations were equipped with Mark 5 digital recording
systems. All correlators were committed to handling the IVS data with
priority processing for meeting timely product delivery
requirements. High capacity disks were purchased and organized in a
common pool to replace magnetic tapes and to obtain additional
recording media capacity. The progress in communication technologies
supported the breakthrough for e-VLBI.  Several tests have been
conducted on national, continental and global levels. Some station are
already connected to fast internet links and regular applications for
e-VLBI (real time or near real time) will be established. The 1-hr
INTENSIVE observation sessions are routinely transferred electronically
to a correlator and will soon be operational. The increased amount of
VLBI data to be produced under the new observing program required
Analysis Centers to handle a larger load. Partially automated analysis
procedures helped to improve the timeliness of product delivery.  \ As
the official IVS product, a complete set of Earth Orientation Parameter
is regularly submitted to the International Earth Rotation Service
(IERS). The set is obtained as a combination of the individual
solutions of the six IVS Analysis Centers (Nothnagel, Steinforth
\cite{nothn01}). Up to the end of the year 2001, the parameters were
derived from the NEOS observations, while since January 2002 the IVS R1
and IVS R4 were used. It should be noted that up to the end of 2001
NEOS was the rapid turn around program which since January 2002 now
includes the IVS R1 and R4 sessions. The objective of the rapid turn
around observation sessions was to minimize the delay between the
observations and the availability of the results.  For the NEOS the
delay was approximately 2 weeks, which has not changed by its
transition to the IVS R4 as all the routine procedures were already
established. For the IVS R4, as new stations were added the data
shipping procedures needed to be set up to be routine. Initially this
caused unexpected delays, which were overcome with time. For the IVS R1
processing, experience needed to be gained at the correlators at Bonn,
Haystack and Washington during the first months. The delay between the
observation to the results is approximately two weeks since April
2002. This should be regarded as significant and real progress, even
though the WG2 goal of only 4 days has not been achieved.  Improvements
for data transmission and a higher throughput at the correlator were
achieved, after the implementation of the newly developed Mark 5
digital data recording system, which has e-VLBI capabilities, allowing
data transmission via high speed Internet links. The determination of
DUT1 from the near-daily 1-hour observations known as "Intensives" have
been carried out since 1983 via the baseline Wettzell-Westford, since
1994 via the baseline Wettzell-Green Bank and since 2000 via the
baseline Wettzell-Kokee Park. These baselines now employ Mark 5 systems
. In the year 2002 a time series observed on the baseline
Wettzell-Tsukba/Japan with the Japanese K4 system was set up. The
regular application of fast Internet links for the INTENSIVE's will
start soon, which will allow rapid availability of DUT1.  \ The VLBI
observations from the IVS R1 and R4 allow determination of tropospheric
parameters, in particular the wet zenith path delay. Since July 2003
the zenith wet path delay is an official IVS product. The University of
Vienna is combining the solutions of five Analysis Centers (Schuh,
Boehm \cite{schuhboehm02}).


\section{Plans}

A "Vision Paper 2010" for geodetic VLBI is under development by
IVS. Considering our increasing requirements (e.g.  the IGGOS project,
the increase in radio frequency interference, our aging antennas)
general refinements and upgrades of VLBI technology are needed in the
future. A Working Group, WG3, was established with the objective to
develop future visions. Goals include unattended observation, improved
global coverage of the network, employment of the new data transmission
technologies and provision of near real time correlation and
products. In collaboration with radio astronomers some guidelines for
future developments will be derived.

\end{document}

%% file: page.tex
\setcounter{page}{309}